\documentclass[runningheads]{llncs}

\usepackage{eccv}

\usepackage{eccvabbrv}

\usepackage{listings}
\usepackage{multirow,array}
\usepackage{graphicx}
\usepackage{booktabs}
\usepackage[percent]{overpic}
\usepackage{multicol}
\usepackage{caption}
\usepackage{subcaption}

\def\ie{\emph{i.e}\onedot} 
 
\def\etc{\emph{etc}\onedot}

\newcommand{\denoiSplit}{\mbox{$\text{denoi}\mathbb{S}\text{plit}$}\xspace}
\newcommand{\microSSIM}{\mbox{$\mathbb{M}\text{icroSSIM}$}\xspace}

\newcommand{\MicroMSSSIM}{\mbox{$\mathbb{M}\text{icroMS}3\text{IM}$}\xspace}

\newcommand\figComarisonWithBaselines{
\begin{figure}[tbh]
\centering
    \includegraphics[width=\textwidth]{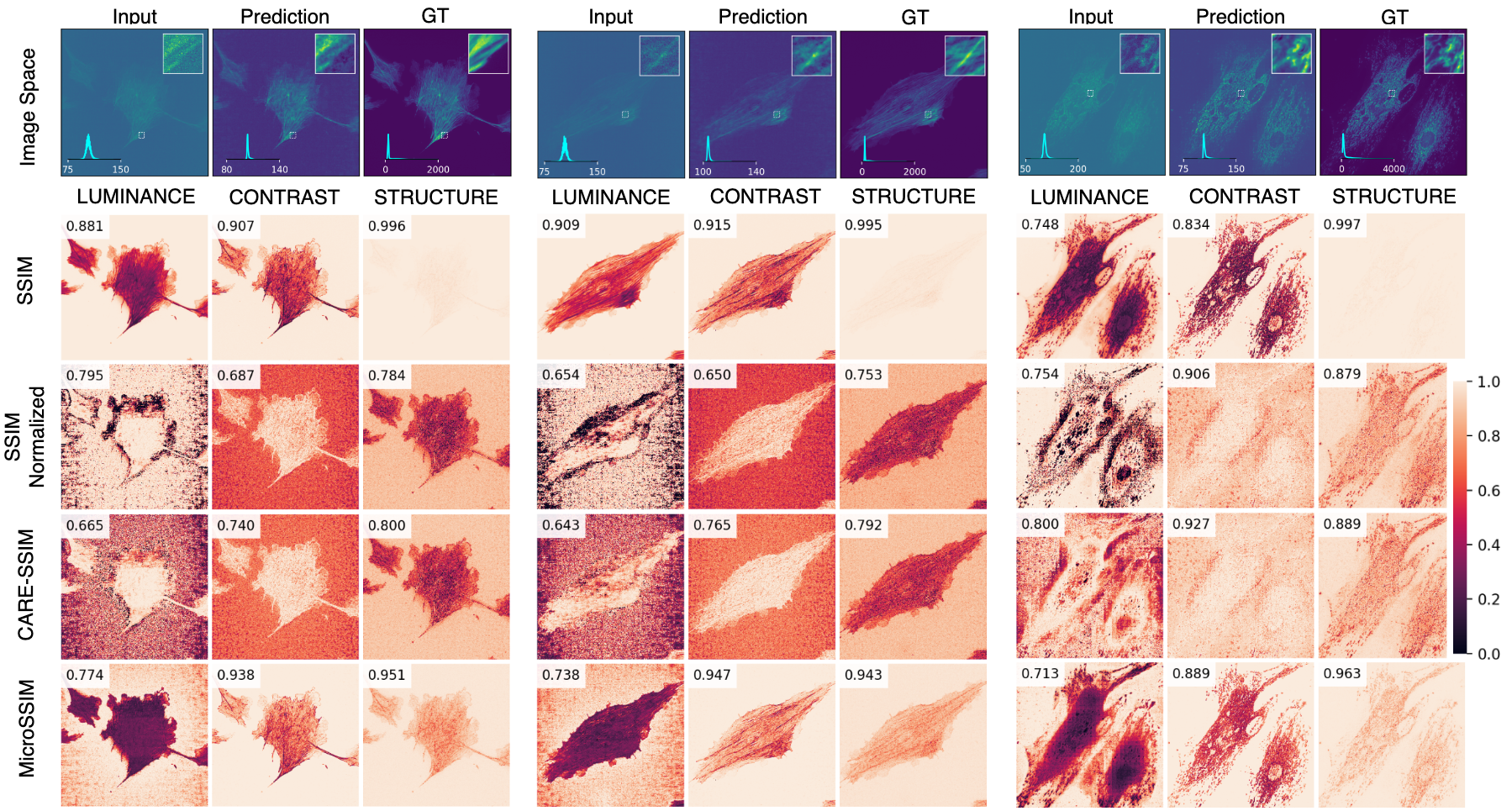}
\caption{\textbf{Comparison with baselines:} In this figure, different SSIM variants are applied between GT and prediction by N2V and we show the three SSIM components namely Luminance, Contrast, and Structure. 
Noisy input, prediction, and noise-free ground-truth comprise the first row with histograms in the inset showing the distribution of pixel intensities. 
We show three examples from Actin and Mitocondria channels of Hagen et al. dataset. 
We compare \textit{SSIM}, vanilla SSIM on unnormalized images (row 2), \textit{SSIM Normalized}, SSIM on mean-std based normalized images (row 3), \textit{CARE-SSIM}, SSIM with scaling and normalization as proposed in CARE~\cite{Weigert2018-pi} (row 4) and \textit{MicroSSIM} (row 5). 
Firstly, the structure component of SSIM is saturated even though the prediction is visibly different from GT. 
Secondly, SSIM-Normalized and CARE-SSIM are highly sensitive to background regions. This does not align with our expectation that the metric should show poor performance if the foreground content is dissimilar and should be less sensitive to background regions. MicroSSIM, as can be seen below, does not seem to suffer from either of the two issues. } 
\label{fig:baseline_comparisons}
\end{figure}
}

\newcommand\figLuminanceAblation{
\begin{figure}[tbh]
\centering
    \includegraphics[width=.98\textwidth]{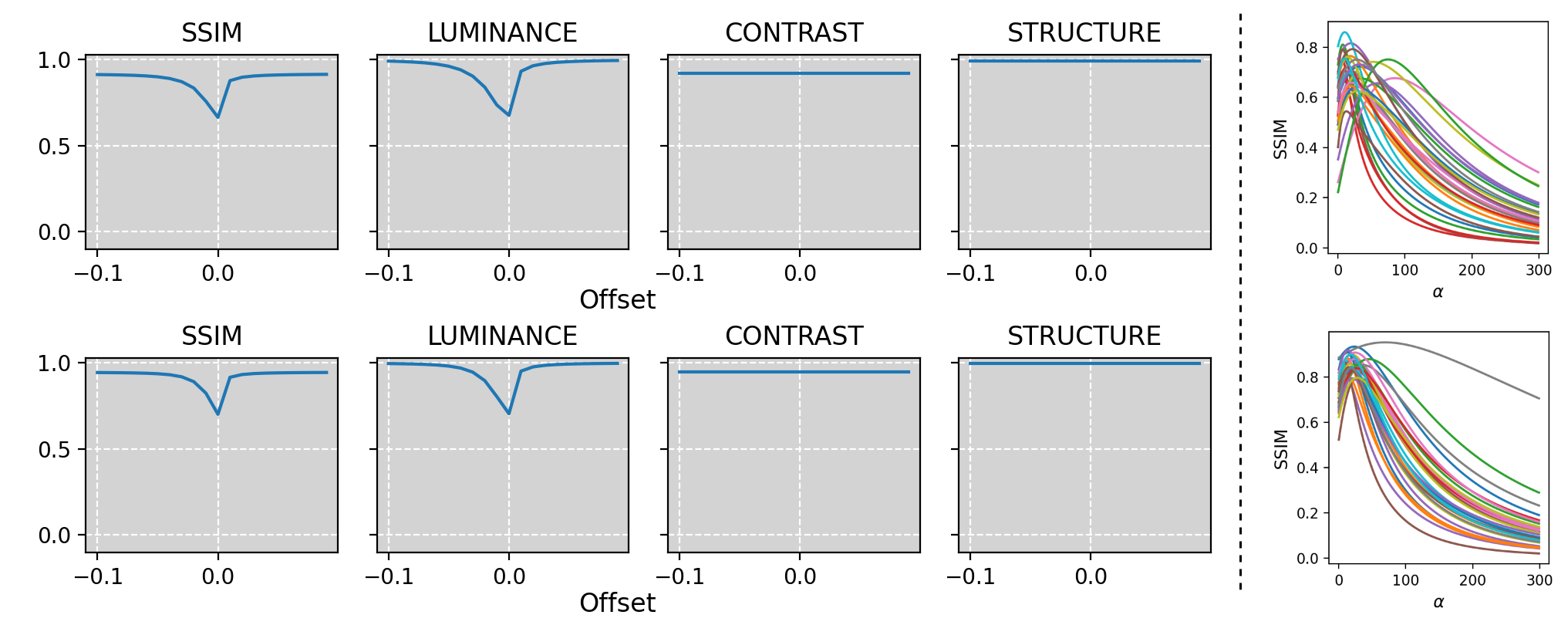}
\caption{\textbf{(Left: Role of offset in luminance)} In this ablation, after doing our proposed normalization of GT and prediction, we add an offset to both GT and prediction and plot the mean SSIM and mean of its components. As can be inferred from the two plots made using two random full frames from Actin (row 1) and Mito (row 2) subdatasets, offset indeed influences the luminance, and therefore, the M-SSIM value. This means that the detector offset set by the microscopist will play a role in SSIM computation. Ideally, only denoising model's performance should play a role. We ensure this by removing the background in our normalization. \textbf{(Right:Uniqueness of $\alpha$)} Here, we took 30 random full frames from Actin and Mito subdataset. Using the pre-trained N2V model, we obtained the denoised predictions and used MicroSSIM to evaluate the prediction. Instead of estimating a single value of $\alpha$ for all images, we manually tried different $\alpha$ values and computed SSIM using it. One can see that for every curve, a unique $\alpha$ exists which gives the maximal SSIM.
}
\label{fig:luminance_and_uniqueness_ablation}
\end{figure}
}

\newcommand\figAblationSaturationActin{
\begin{figure}[]
\centering
    \includegraphics[width=.90\textwidth]{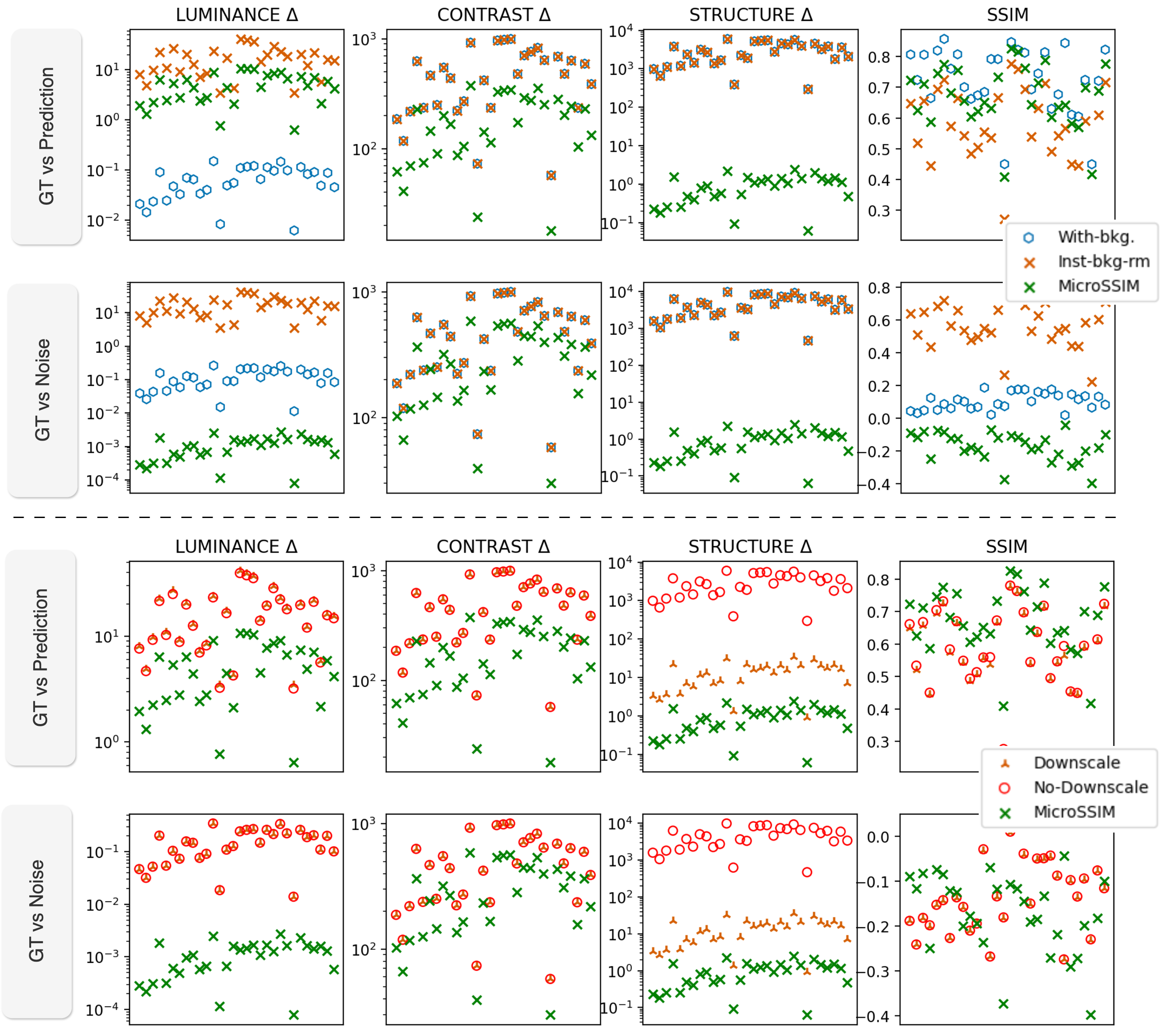}
\caption{\textbf{Inspecting role of pre-processing on saturation factor $\Delta$ for Actin denoising task:} 
The top and bottom panels investigate the role of background and downscaling respectively. 
All baselines used here do not use $\alpha$. 
Only \microSSIM uses it.
Lower $\Delta$ values indicate higher sensitivity of SSIM towards the (dis)similarity between GT and prediction. 
So, lower values are preferable.  
In the second row of both panels, we compare GT with a purely noisy image drawn from a uniform distribution.
In this case, lower SSIM values are better. 
\textbf{(Top)} 
\textit{With-Bkg} baseline does not remove the background in the pre-processing step, and \textit{Inst-bkg-rm} baseline first estimates the background from the given GT and prediction image separately and then removes it from the two images.
Note that in \microSSIM, background estimation does not happen separately for a single image, but happens once at the dataset level.
\textbf{(Bottom)} \textit{Downscale} and \textit{No-Downscale} baselines , similar to \microSSIM, removes the background in the pre-processing step. 
\textit{Downscale} also divides the GT and prediction with the maximum pixel value present in the GT but \textit{No-Downscale} does not.
}
\label{fig:saturation_ablation_actin}
\end{figure}
}

\newcommand\figAblationSaturationMito{
\begin{figure}[]
\centering
    \includegraphics[width=.95\textwidth]{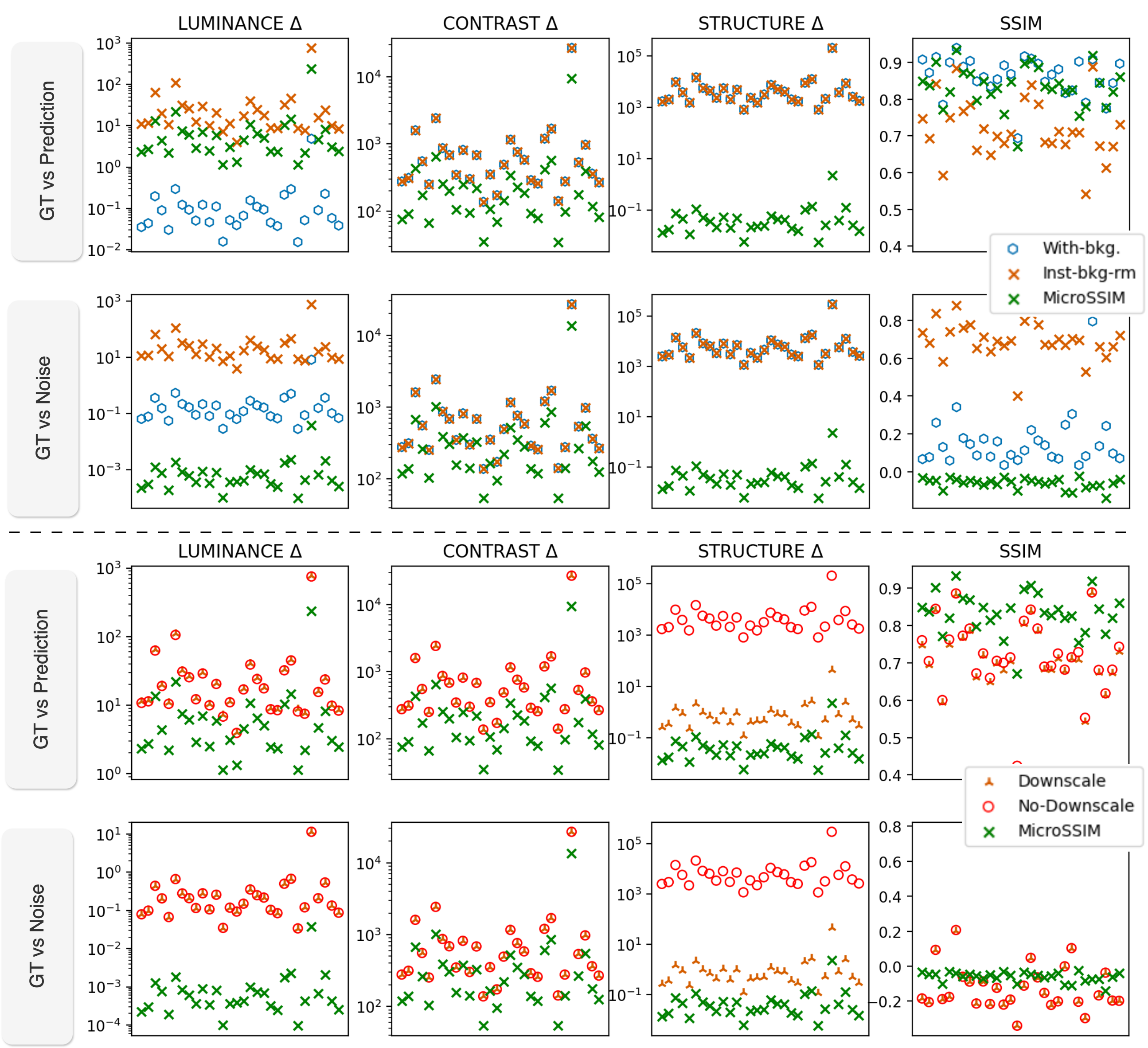}
\caption{\textbf{Inspecting role of pre-processing on saturation factor $\Delta$ for Mitochondria denoising task:} 
The top and bottom panels investigate the role of background and downscaling respectively. 
All baselines used here do not use $\alpha$. 
Only \microSSIM uses it.
Lower $\Delta$ values indicate higher sensitivity of SSIM towards the (dis)similarity between GT and prediction. 
So, lower values are preferable.  
In the second row of both panels, we compare GT with a purely noisy image drawn from a uniform distribution.
In this case, lower SSIM values are better. 
\textbf{(Top)} 
\textit{With-Bkg} baseline does not remove the background in the pre-processing step, and \textit{Inst-bkg-rm} baseline first estimates the background from the given GT and prediction image separately and then removes it from the two images.
Note that in \microSSIM, background estimation does not happen separately for a single image, but happens once at the dataset level.
\textbf{(Bottom)} \textit{Downscale} and \textit{No-Downscale} baselines , similar to \microSSIM, removes the background in the pre-processing step. 
\textit{Downscale} also divides the GT and prediction with the maximum pixel value present in the GT but \textit{No-Downscale} does not.
}
\label{fig:saturation_ablation_mito}
\end{figure}
}

\newcommand\figBkgAblationMicroSSIM{
\begin{figure}[tbh]
\centering
\includegraphics[width=.8\textwidth]{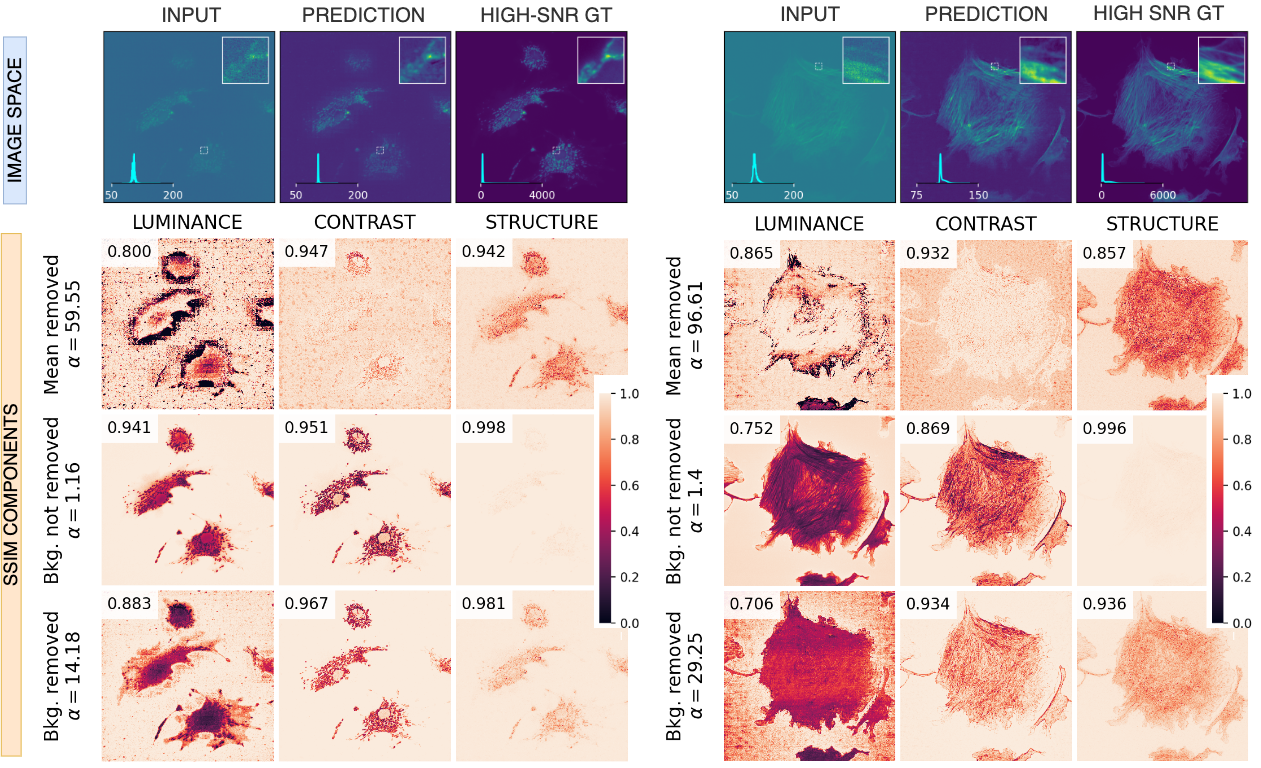}
\caption{\textbf{Background removal ablation:} 
We show SSIM components and estimated $\alpha$ with background removal disabled (row 3), enabled (row 4), and mean removal (row 2).  
We use two random full frames from Actin and Mito sub-datasets. 
Without background removal, $\alpha$ gets underestimated.
With mean removal, $\alpha$ gets overestimated.
}
\label{fig:bkg_removal_ablation}
\end{figure}
}

\newcommand\figTeaser{
\begin{figure}[tbh]
\centering
\includegraphics[width=.9\textwidth]{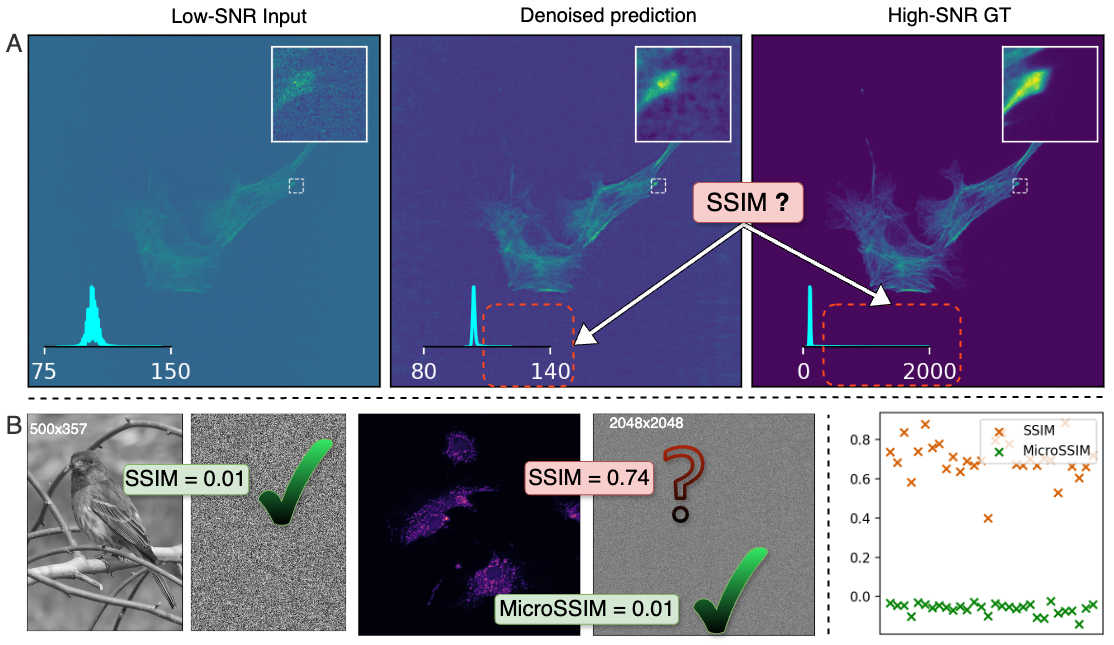}
\caption{\textbf{Failure mode of SSIM on Microscopy Data.} 
\textbf{(A)}  A noisy microscopy image, \ie  a micrograph, its denoised version predicted using N2V, and the corresponding High-SNR (noise free) ground truth is shown. 
There is a problem in the evaluation of denoising quality, which is that the pixel intensity distribution of the ground truth and the prediction (as shown in respective insets) are very different. 
This is specifically true for the foreground content which comprises brighter pixels. 
So applying SSIM directly on it will not give a sensible value. 
We solve this issue with \microSSIM.
\textbf{(B)} We show one example to demonstrate an apparent counter-intuitive behavior of SSIM. The SSIM between a natural image (taken from Imagenet~\cite{imagenet}) and a pure noisy image drawn from the uniform distribution is much lower than the SSIM between a micrograph and a noisy image with identical distribution as before.
The expectation naturally is to have $\text{SSIM}\approx 0$ in both cases.
We solve this issue with \microSSIM and appropriate data pre-processing and show it in bottom right plot where 30 random microscopy images are used.
}
\label{fig:teaser}
\end{figure}
}

\newcommand\figMSSSIMissues{
\begin{figure}[!tbp]
\centering
  \begin{subfigure}[b]{0.3\textwidth}
    \includegraphics[width=\textwidth]{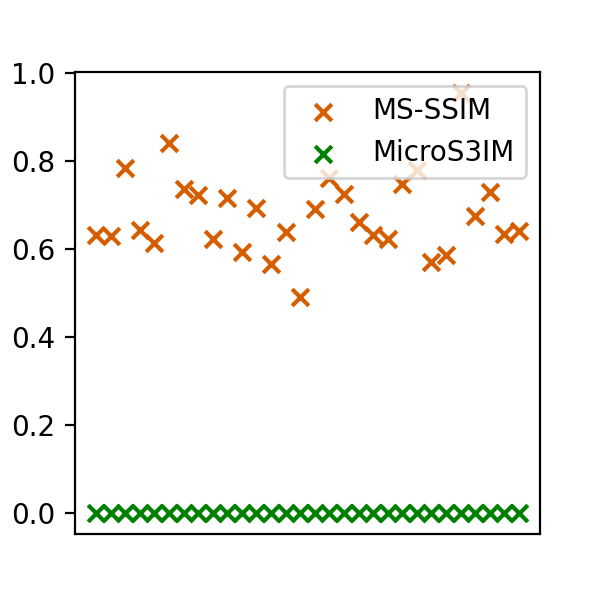}
  \end{subfigure}
  \begin{subfigure}[b]{0.3\textwidth}
    \includegraphics[width=\textwidth]{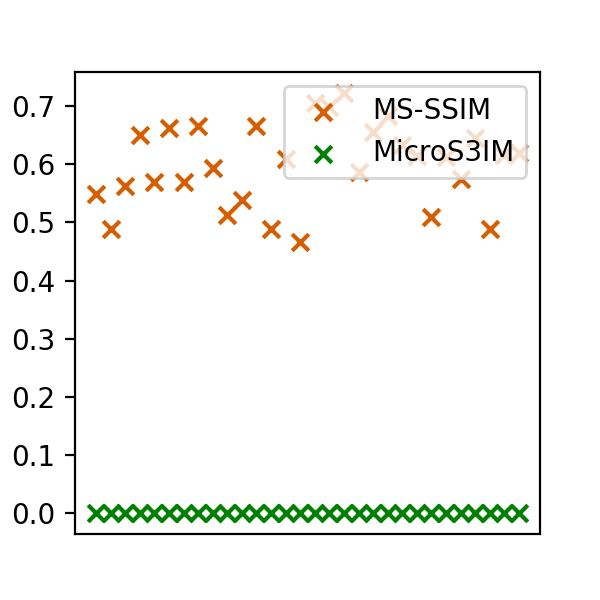}
  \end{subfigure}
  \caption{\textbf{Issues with Multiscale-SSIM:} Similar to SSIM, with MS-SSIM as well, we see the similar behavior that comparison of unnormalized micrographs with pure noisy images yields a large score indicating the presence of large saturation. With our \MicroMSSSIM, where we have the pre-processing step and subsequent scaling of the prediction with $\alpha$, we get the desired result, \ie, values very close to zero.}
  \label{fig:msssimIssues}
\end{figure}
}

\usepackage[accsupp]{axessibility}  

\usepackage{hyperref}

\usepackage{orcidlink}
\usepackage{amsmath}
\usepackage{amsmath}
\DeclareMathOperator*{\argmax}{arg\,max}
\DeclareMathOperator*{\argmin}{arg\,min}
\newcommand{\supsection}[1]{Supp.\ \textit{Sec.\ #1}}
\newcommand{\supfigure}[1]{\textit{Supp.\ Fig.\ #1}}
\newcommand{\suptable}[1]{\textit{Supp.\ Tab.\ #1}}

\begin{document}

\title{MicroSSIM: Improved Structural Similarity for Comparing Microscopy Data} 

\titlerunning{\microSSIM}

\author{Ashesh Ashesh\orcidlink{0000-0003-3778-0576} \and
Joran Deschamps\orcidlink{0000-0001-8462-2883} \and
Florian Jug\orcidlink{0000-0002-8499-5812}}

\authorrunning{Ashesh, F.\ Jug}

\institute{
Fondazione Human Technopole, Viale Rita Levi-Montalcini~1, 20157 Milan, Italy
\email{ashesh276@gmail.com, florian.jug@fht.org}}

\maketitle

\begin{abstract}
  Microscopy is routinely used to image biological structures of interest.
  Due to imaging constraints, acquired images, also called as micrographs, are typically low-SNR and contain noise. 
  Over the last few years, regression-based tasks like unsupervised denoising and splitting have found utility in working with such noisy micrographs.
  For evaluation, Structural Similarity (SSIM) is one of the most popular measures used in the field.
  For such tasks, the best evaluation would be when both low-SNR noisy images and corresponding high-SNR clean images are obtained directly from a microscope. 
  However, due to the following three peculiar properties of the microscopy data, we observe that SSIM is not well suited to this data regime:
  (a) high-SNR micrographs have higher intensity pixels as compared to low SNR micrographs, (b) high-SNR micrographs have higher intensity pixels than found in natural images, images for which SSIM was developed, and (c) a digitally configurable offset is added by the detector present inside the microscope which affects the SSIM value.
  We show that SSIM components behave unexpectedly when the prediction generated from low-SNR input is compared with the corresponding high-SNR data.
  We explain this behavior by introducing the phenomenon of saturation, where the value of SSIM components becomes less sensitive to (dis)similarity between the images. 
  We propose an intuitive way to quantify this, which explains the observed SSIM behavior.
  We introduce \microSSIM, a variant of SSIM, which overcomes the above-discussed issues. 
  We justify the soundness and utility of \microSSIM using theoretical and empirical arguments and show the utility of \microSSIM on two tasks: unsupervised denoising and joint image splitting with unsupervised denoising.  
  Since our formulation can be applied to a broad family of SSIM-based measures, we also introduce \MicroMSSSIM, a microscopy-specific variation of MS-SSIM. 
  The source code and python package is available at https://github.com/juglab/MicroSSIM. 
\end{abstract}

\figTeaser
\section{Introduction}
\label{sec:intro}
Microscopy is routinely used to image structures like proteins, cell organelles, cytoskeletal components, etc. 
Due to numerous imaging constraints, like having faster acquisition, preventing photo-bleaching, limiting radiation damage to biological samples, \etc, microscopists typically work with configurations that generate relatively low-SNR micrographs, \ie, images containing noise. 

While denoising is, in itself useful, over the years it has been integrated into other classification pipelines like segmentation, tracking, and more recently with regression-based tasks like image splitting~\cite{denoisplit}. 
With the advent of deep learning, several supervised and self-supervised denoising methods~\cite{n2v-krull, Prakash2021-dz} have been developed.
Our focus is on arguably the most practically useful family of denoising methods, namely unsupervised denoising, where high-SNR data is not needed for training the denoiser model.

Given the utility and popularity of unsupervised denoising, it becomes important to have a reliable way to evaluate the denoising performance of the different prevalent approaches. 
To do that, one needs to have (a) an evaluation dataset consisting of tuples of low-SNR input images and corresponding high-SNR ground-truth images and (b) to have a reliable measure. 
Our work focuses on the latter part, towards creating a reliable measure for evaluation of denoising performance on microscopy data. 
Before we embark on that journey, in the next paragraph, we outline the evaluation dataset category we cater to. 

In several existing methods that perform unsupervised denoising~\cite{denoisplit,n2v2}, authors synthetically add noise to existing clean datasets. 
In these setups, they trivially have pairs of noisy and clean data.
However, the limitation is that the noise is synthetic.
Other approaches that do not have the above-mentioned limitation fix a field of view (FOV) and take multiple noisy snapshots from the microscope. 
Here, the noise is naturally real.
Averaging the noisy images taken on the same FOV, which means having the same underlying content, leads to the corresponding higher SNR image~\cite{Weigert2018-pi}.
In this case, the limitation, although less severe, is that the high SNR image is synthetic.
A more rigorous way exists wherein the aggregation is not done outside the microscope as discussed above, but rather, the aggregation happens inside the microscope itself. 
Specifically, a field of view is fixed and the noisy input is imaged. 
Subsequently, the acquisition time and/or the laser power is increased and another acquisition is made. 
This yields the corresponding high SNR image. 
The focus of this work is to develop a measure that is to be used on the evaluation dataset generated by the last, arguably the most rigorous method.

Next, we focus our attention on evaluation measures. 
In the computer vision community, Structural similarity (SSIM)~\cite{ssim_paper} is one of the most popular measures used on regression tasks and therefore it is no surprise that most of the denoising methods report their performance using SSIM. 
SSIM constitutes a combination of three quantifiable terms which quantifies luminance, contrast, and structure.
Over the last decade, many variants of SSIM have been introduced, each improving upon some specific aspects of the original measure. 
This work focuses on adapting the SSIM-based measures for microscopy data.  

In this work, we observe that there are three peculiar properties of the microscopy data that make SSIM and its variants, when used in its default way, an inappropriate evaluation measure for the evaluation dataset type in focus. 

Firstly, high-SNR micrographs occupy a larger portion of the dynamic range. 
In simple terms, pixel values in high-SNR images are higher on average than what is found in low-SNR micrographs. 
Note that an unsupervised model, trained solely on low-SNR data, naturally yields predictions where the average pixel intensity of a predicted patch matches that of the corresponding noisy low-SNR input patch. 
This is a problem because the corresponding high-SNR ground truth will have much higher pixel intensities on average and so, even though they contain the same structure, SSIM will behave unexpectedly.
We depict this problem in Fig.~\ref{fig:teaser}.
Concisely put, the difference in the pixel intensity levels between high-SNR and low-SNR micrographs makes SSIM inappropriate in this data regime.

The other peculiar property of the microscopy data is that the microscope's detector adds a configurable scalar offset to all pixels. Background regions in the micrograph, regions where there is no structure typically have pixel values around this offset. 
This can lead to two potential problems. Firstly, it is possible that the offsets present in the high-SNR and the low-SNR micrographs are different. 
In such a situation, even if the denoiser perfectly denoises the input, the pixel values will, on average, differ by a constant offset which will wrongly degrade the luminance computation.
Secondly, even if the offset is identical in both low-SNR and high-SNR images, its value changes the luminance component. 
We show this effect in the experiments section. 

Thirdly, as SSIM was originally developed for and still is mostly used with natural images, we have discovered that SSIM has issues when working with images with a much higher range of pixel intensities, as is the case with high-SNR micrographs.
Specifically, SSIM gives more importance to constant hyperparameters originally put in the SSIM expression for stability and less importance to the similarity/dissimilarity between the two input images (see Fig.~\ref{fig:teaser}).
In this work, we denote this effect as saturation.
Our solution to these problems is to apply a suitable pre-processing step on both the target and prediction and subsequently apply a linear transformation on the prediction.

But before we present our solution, it is worthwhile to ponder on the desirable properties of a useful measure. 
We argue that the measure should not focus too much on the background regions. 
This is because the primary objective of microscopy data is to image the foreground content.
The background is present only because it is inevitable to not have it.
It is worth contrasting this with natural images where sky, sea, or other typical background content is very much relevant to the image.
At the same time, if the predictive model introduces artifacts in the background, then that needs to be accounted for. 
So, segmenting out the background region does not look promising even when assuming perfect segmentation, which itself is challenging with complicated structures.
With regards to linear transformation, it should be noted that a sub-optimal transformation could easily introduce a fictitious gap between the backgrounds of the target and prediction.
As we show in the experiments section, one of the popular methods applying a linear transformation, CARE-SSIM~\cite{Weigert2018-pi}, suffers from this issue.

To handle these microscopy data peculiarities, we introduce \microSSIM, a variant of SSIM measure. 
We have designed \microSSIM such that it can handle the intensity level difference between the two images. 
So, with \microSSIM, we can now evaluate the performance of a denoiser's output predicted from a low-SNR image against a high-SNR ground truth. 
More specifically, we infer a scalar $\alpha$ which we multiply with the prediction before computing the SSIM. 

However, for \microSSIM to be effective, we also propose a pre-processing step where we remove the estimate of the offset (added by the detector) $\beta$ from both the target and the predicted image. 
We additionally have a downscaling step, wherein we divide the resultant images with a scalar. 
We show that the downscaling step limits saturation. 

We show both theoretically and empirically the uniqueness of the scalar $\alpha$ which yields globally optimal SSIM value. 
This uniqueness allows us the freedom to use an off-the-shelf optimization module to estimate a globally optimal $\alpha$. 
However, we note that \microSSIM is not designed to yield higher values.
Rather it is designed to learn a linear mapping between sets of predictions done on low-SNR images and the corresponding high-SNR images. 
Therefore we estimate a single $\beta$ and $\alpha$ over the whole dataset.

Finally, our implementation is such that it is easily extendable to other SSIM variants. 
To this end, we extend Multiscale SSIM (MS-SSIM)~\cite{ms_ssim} to \MicroMSSSIM.

Concisely put, in this manuscript, we report the incompatibility of a family of SSIM variants to microscopy data on multiple fronts which we show quantitatively. 
We introduce a quantifiable concept of saturation to explain the issues faced by SSIM variants.
We modify the SSIM measure and propose a new measure, \microSSIM which resolves these issues. 
We show the sensibility of our measure using both theoretical and empirical evidence. 
Our approach can easily be extended to other measures of the SSIM family and we show that by extending MS-SSIM to \MicroMSSSIM.
We show the utility of our approach on two practical tasks namely unsupervised denoising and joint splitting with unsupervised denoising. 
\section{Related work}
\label{sec:relatedwork}
The origin of the Structural Similarity measure (SSIM) could be traced back to the universal quality index (UQI)~\cite{universal_image_quality}. In this seminal work, the image similarity was decomposed into three measurable quantities namely luminance($l$), contrast ($c$), and structure($st$). 
Later, constants $c_1,c_2,c_3$, which depended on the range of pixel values, were added to each of the three measurable quantities proposed in UQI formulation to avoid division by zero~\cite{ssim_paper}. When computing SSIM between two images $x$ and $y$, individual terms had the following expression.

\begin{minipage}{.24\textwidth}
\begin{equation*}
  \text{l} = \frac{c_{1} + 2 u_{x} u_{y}}{c_{1} + u_{x}^{2} + u_{y}^{2}},
\end{equation*}
\end{minipage}%
\begin{minipage}{.21\textwidth}
\begin{equation*}
  \text{c} = \frac{c_2 + 2s_x s_y}{c_2 + s_x^2 + s_y^2}\text{ ,}
\end{equation*}
\end{minipage}
\begin{minipage}{.21\textwidth}
\begin{equation*}
  \text{st} = \frac{c_3 + s_{xy}}{c_3 + s_x s_y}\text{ ,}
\end{equation*}
\end{minipage}
\begin{minipage}{.29\textwidth}
\begin{equation}
  \text{SSIM} =  l^{\alpha}c^{\beta}{st}^{\gamma}
\end{equation}
\end{minipage}\break

where $u_{z}=\sum_k^K \frac{z[k]}{K}$, $s_z = \sqrt{\sum_k^K \frac{(z[k] - \mu_z)^2}{K-1}}$ and $s_{x,y} = \frac{\sum_k^K (x[k]-\mu_k)(y[k] -\mu_k)}{K-1}$.
Here, $z \in \{x,y\}$.
For mathematical convenience, $c_3=c_2/2, \alpha=\beta=\gamma=1.0$ was introduced which enabled simplification of the expression and it is this formulation which people refer to as SSIM today. 
Mathematically, the SSIM measure between two images $x$ and $y$ is computed as follows:
\begin{equation}
    \text{SSIM}(x,y) = \frac{\left(c_{1} + 2 u_{x} u_{y}\right) \left(c_{2} + 2 s_{xy}\right)}{\left(c_{1} + u_{x}^{2} + u_{y}^{2}\right) \left(c_{2} + s_{x}^{2} + s_{y}^{2}\right)}
\end{equation}

Typically, SSIM is not computed over the entire image but rather over small overlapping regions in an efficient way using convolution operation.
The result is that one has an SSIM value for every pixel of the predicted image which is sufficiently away from the image boundary.
To get a single value for the entire image, MSSIM is used which is the average of all SSIM values computed over all small regions of the image. 
In practice, the MSSIM measure is what researchers refer to even when they use the term SSIM. Technically speaking,
\begin{equation}
    \text{MSSIM(x,y)} = \frac{1}{HW} * \sum_{i,j}\text{SSIM}(x_{i,j},y_{i,j}),
\end{equation}
where H and W are the height and width of the images and $x_{i,j}$ and $ y_{i,j}$ denote the patch from x and y centered around $(i,j)$ location respectively.

As mentioned before, since SSIM was computed on small patches, there was a need to design a variant that could compare larger structures. 
Multiscale SSIM (MS-SSIM)~\cite{ms_ssim} was developed for this purpose. 
In MS-SSIM, one creates a sequence of length five of successively downsampled and low-pass filtered versions for both images. 
The contrast and structure components are computed at each of these five levels and the luminance is computed at the last level. 

Since then, there has been the development of numerous variants of SSIM, each catering to a specific purpose. Complex wavelet SSIM (CW-SSIM)~\cite{cw_ssim} was developed to have less penalty for small rotations and translations. 3D-SSIM~\cite{3d_ssim} was developed for videos. Spherical SSIM~\cite{spherical_ssim} was developed to work well on spherical projections.   

Weigert et al.~\cite{Weigert2018-pi} introduced an SSIM variant which we refer to as CARE-SSIM. 
It tried to handle the issue of mismatch in the intensity scales between high-SNR ground truth (GT) and predictions done on low-SNR input by learning a linear transformation of the prediction which minimized the MSE loss with high SNR GT. 
Later works~\cite{Prakash2021-dz, Ashesh2023-wtf,denoisplit} converted the target and prediction to zero mean and learned only the scaling factor (scalar) to be multiplied with the prediction with the same objective of minimizing MSE. 

\section{Our approach}
\label{sec:ourapproach}
We begin this section by asking the question: what are the desirable properties of a measure operating on microscopy data? 
For us, it is that the measure should focus on the foreground region and it should be sensitive to the inputs being fed, \ie, if the prediction is considerably worse, the score should also be unambiguously low. 
With this in mind, we quantify the aspect of sensitivity towards the input images by formulating the quantifiable phenomenon of saturation. 
Next, we propose the pre-processing scheme that handles the detector-offset issue and the issue of large intensities. 
Lastly, we describe the methodology to estimate the linear scalar $\alpha$ that gets multiplied with the prediction. 
\subsection{Saturation}
\label{subsec:saturation}
In this subsection, we explore one of the aspects of SSIM that becomes an issue when images have large pixel intensities, specifically when the difference between maximum and minimum pixel intensity in the ground truth image, henceforth abbreviated as $\gamma$, is large. 
Each of the three SSIM components has the following general expression 
\begin{equation}
g_i = \frac{a+c_i}{b+c_i},    
\end{equation}
where $\{c_i, i\in\{1,2, 3\}\}$, as defined at Sec.~\ref{sec:relatedwork}, were introduced for stability. 
All $c_i$ values are in turn computed from the $\gamma$ and a set of hyper-parameters $k_i, i\in\{1,2,3\}$ as $c_i = (k_i \times \gamma)^2$.
Now, as $\gamma$ becomes large, so does $c_i$. 
The problem arises when $a,b$ do not, for some reason, become equally large. 
To drive home the point, one could observe the following extreme situation: given $a,b\in R$, $\lim_{c_i\to\infty} g = 1$, irrespective of what $a,b$ are, \ie, irrespective of how similar the two images are. 
We coin the term \textbf{saturation} to denote this aspect of SSIM components where the value becomes less sensitive to the (dis)similarity between the two images. 
We design the following expression to quantify it: 
\begin{equation}
    \Delta = min(|\frac{c_i}{a}|, |\frac{c_i}{b}|)
\end{equation}

The larger the $\Delta$ is, the more the saturation.
So, when $\Delta$ is large for an SSIM variant measure, that would mean a larger measure value and at the same time less sensitive measure.  
It is important to note that a large $\Delta$ is not incorrect: the SSIM component will still monotonically increase to 1 as $a$ and $b$ become more and more similar.
However, the sensitivity can be significantly affected, as we show in the experiments section.  
Since $\Delta$ is estimated for each pixel, we average it over the pixels to get a single value when required. 
In Sec.~\ref{sec:experiments}, we show that \microSSIM attains lower $\Delta$ than other approaches.
\subsection{Handling background and large pixel intensities}
In Microscopy data, a digital offset is added by the detector present inside the microscope. 
This offset value, modulo the Gaussian noise, is what is typically observed in the background regions of the images. 
Offsets are added in both low-SNR and high-SNR acquisitions and they are relatively similar.
However, the foreground pixels of high-SNR images have much higher pixel values than those of low-SNR images.
Hence, scaling the prediction with the objective to make foreground pixels in both data regimes to have similar values will adversely scale up background pixel values. 
This will be undesirable since the background values of the low-SNR scaled image will become higher than those of high-SNR GT. 

To handle the offset issue, we do a pre-processing step and do background subtraction. 
We compute the background offset $\beta$ separately for the GT and prediction. 
Micrographs often have the lowest pixel value well below the average background pixel value.
Using the minimum value would therefore be an incorrect estimate for background. 
So, we assume $3^{rd}$ percentile of the pixel intensities to be the background.
We compute the percentile not on individual images, but rather on the whole dataset. 
Since background regions typically occupy more than three percent of all pixels in a micrograph, we find our empirical choice reasonable since the value changes slowly around this percentile.
For instance, $0^{th}$, $1^{st}$, $3^{rd}$ and $9^{th}$ percentile for Actin GT micrographs used in this study are 41, 100, 102, and 105 respectively. 
However, for more densely populated micrographs, a lower percentile might be optimal.
Alternatively, one can manually inspect a few micrographs and estimate an average pixel value for the background.
While all of these are reasonable options, what is important to note is that when comparing multiple predictors on any dataset, the same offset must be used for GT.
This is because the SSIM score depends on the offset, which we show in Sec.~\ref{sec:experiments}. 

Next, high-SNR micrographs have much larger pixel intensities.
We show in Sec.~\ref{sec:experiments} that it leads to a higher saturation of the structure component of the SSIM. 
To fix this, we divide the prediction and GT with the maximum pixel intensity found in GT images.

\subsection{Optimization to obtain the scaling factor $\alpha$}
In this section, we describe our approach to obtain $\alpha$, a positive real number with which we scale our pre-processed prediction.
A plausible objective could be to obtain a scalar $\alpha$ such that $\alpha = \argmax_{\alpha} \text{SSIM}(x, \alpha y)$. 
We show the expression of SSIM operating on scaled prediction $\alpha x$ and GT $y$ below.
\begin{equation}
SSIM(x, \alpha y) = \frac{\left(2 \alpha u_{x} u_{y} + c_{1} \right) \left(2 \alpha s_{xy} + c_{2}\right)}{\left(\alpha^{2} s_{y}^{2} + c_{2} + s_{x}^{2}\right) \left(\alpha^{2} u_{y}^{2} + c_{1} + u_{x}^{2}\right)}
\label{eq:alphaSSIM}
\end{equation}
The analytic way would be to compute the derivative of $\text{SSIM}(x, \alpha y)$ with respect to $\alpha$ and set it to zero to obtain the extrema solutions. 
The sign of the second derivative along with the constraint that $\alpha >= 0$ could then be used to pick the desired maxima.

In \supsection{\ref{sec:deriveoptimalalpha}}, we show the full expression of the derivative of SSIM, which is a higher-order polynomial for which the existing equation solvers were not able to give a closed-form solution. 
However, under the assumption that $c_1 = c_2 = 0$, we prove in \supsection{\ref{sec:deriveoptimalalpha}} that the following unique closed-form solution for the maxima exists  
\begin{equation}
    \alpha = \sqrt{\frac{s_{x} u_{x}}{s_{y} u_{y}}}.
\label{eq:optimalalpha}
\end{equation}
Moreover, in Sec.~\ref{sec:experiments}, we empirically show over multiple denoising tasks that a unique maxima exists in the general case as well. 

The presence of a global optimum allowed us to use \textit{minimize} function from \textit{scipy.optimize} package, an off-the-shelf optimizer, to minimize the following expression $\alpha = \argmin_{\alpha} \sum_{(x,y)} -1*SSIM(x, \alpha y)$, where SSIM as defined in Equation~\ref{eq:alphaSSIM} is used and $y,x$ are the denoised prediction and the corresponding GT respectively. 
Once $\alpha$ and $\beta s$ are estimated, \microSSIM between a prediction y and GT x is
\begin{equation}
\text{\microSSIM(x,y)} = \text{SSIM}(\frac{x-\beta_{GT}}{max_{GT}}, \frac{\alpha (y - \beta_{pred})}{max_{GT}}).
\end{equation}
Here, $max_{GT}$ is the maximum pixel intensity observed in all GT images. It is important to stress the fact that we learn a single scalar for the entire dataset. 
Had we optimized for every $(x,y)$ pair, we would get a higher measure value on average, but this does not align well with the motivation for this measure, which is to estimate an optimal linear transformation between the space of predictions to their corresponding high-SNR micrographs.
Naturally, this linear transformation should be the same for every GT image and prediction pair in the dataset. 

Since our approach does not alter the logic of the measure itself, it is relatively straightforward to apply our approach to measures other than SSIM. 
To this end, we extend our approach and introduce \MicroMSSSIM, an extension to MS-SSIM (Multiscale SSIM). We use the same pre-processing and $\alpha$ scaling and subsequently compute MS-SSIM between pre-processed high SNR GT and scaled prediction. 

In terms of time, \microSSIM computation takes the same time as SSIM. 
The one-time optimization cost for $\alpha$ is relatively low: it takes less than 1 minute for the optimization when working with 25 frames of size $2048\times2048$ on an AMD EPYC 7763 64-Core processor. 
\figComarisonWithBaselines
\section{Experiments}
\label{sec:experiments}
\paragraph{\textbf{Dataset and training details}}
We selected high-SNR and corresponding low-SNR images of Actin and Mitochondria channels taken from the publicly available Hagen et al~\cite{Hagen2021-xh}. 
For training \denoiSplit, we use the same code, hyper-parameters, and train/val/test split as proposed in~\cite{denoisplit}.
For \denoiSplit, we evaluate the test set. 
For N2V, we evaluate on all 100 frames. We used CAREamics library~\footnote{https://github.com/CAREamics} for training N2V.

\paragraph{\textbf{Baselines}}
We work with two baselines. 
First baseline is the vanilla SSIM measure. As the second baseline, we use the CARE-SSIM measure. 
In the ablation studies, we create several other baselines which we define when they are needed.
\figAblationSaturationMito

%
\paragraph{\textbf{Models}}
For the denoising task, we work with N2V~\cite{n2v-krull}, one of the most cited unsupervised denoising methods. For joint splitting and unsupervised denoising, we work with \denoiSplit~\cite{denoisplit} which is SOTA for this task. 
\paragraph{\textbf{Comparison with baselines}}
In Fig.~\ref{fig:baseline_comparisons}, we compare \microSSIM with the different baselines. 
In \textit{vanilla SSIM} (row 2) baseline, we compute SSIM between the unnormalized GT and prediction. 
One can observe a couple of issues with SSIM. 
We show pixel intensity distributions in the inset of input, prediction, and high-SNR GT (row 1). 
\figBkgAblationMicroSSIM
\figLuminanceAblation

One can observe that the pixel intensity distribution of prediction is at a different scale from that of the high-SNR GT.
Despite this difference, we see quite high SSIM components, which is a matter of concern. 
One of the reasons for this is that for background regions, which form a relatively large portion of the micrographs, the intensity distributions match. 
More specifically, the third percentile was on average $102$ for both GT and the prediction. 
But, even in the foreground regions, the structure component is nearly perfect.
We argue that it is because of the relatively higher saturation of that component. 
This is the case because, despite visual differences between the prediction and the target, the structure component is very close to 1. 
Please refer to the experiment for quantitative evaluation of $\Delta$ for more details. 

Our next baseline is SSIM computed on normalized GT and prediction, with normalization done using their respective mean and standard deviation (row 3).
We observe that the background gets heavily penalized since the normalization made the background pixel intensities different in GT versus the prediction.
This is not preferable since we want our measure to focus on the foreground and ignore the background as long as there are no unwanted structures in the prediction in those regions. 
Similar is the situation with our next baseline, CARE-SSIM (row 4), albeit due to a different reason. 
CARE-SSIM uses MSE to estimate the optimal transformation. 
Since MSE focuses on higher-intensity pixels, the background, which comprises the lowest pixel intensities, suffers from the lack of fitting.
We also argue that when the aim is to work with SSIM, it makes more sense to optimize SSIM itself, which is what \microSSIM does, than optimizing the MSE for getting the transformation.
\microSSIM (row 5) arguably is superior to the baselines on these abovementioned issues.
%
%
\paragraph{\textbf{Relevance of background, downscaling and $\alpha$ in Saturation}}
In Fig.\ref{fig:saturation_ablation_mito}, we do ablations on the Mitochondria denoising task to better understand the factors that affect Saturation. 
A similar ablation using Actin denoising task can be found at \supfigure{\ref{fig:saturation_ablation_actin}}. 
The top panel investigates the role of the background. The bottom panel investigates the role of downscaling, \ie, dividing both prediction and GT by the maximum intensity present in GT. 
For this ablation, we pick 30 random full-frame images from each task. 
We use the trained N2V to get their denoised predictions. 
In the first row of each panel, we compute the average saturation factor $\Delta$ for each of the three SSIM components using the denoised prediction and GT and show it with on a scatter plot.
In the second row, we compare GT with an image containing purely noise drawn from the uniform distribution.
A better measure should yield lower SSIM values in the second row and a lower saturation factor across both rows and SSIM components.

We observe that when the background is not removed, then one typically has higher $\Delta$ and therefore produces a higher SSIM value, which is undesirable. 
We observe that while the downscaling operation does not change SSIM, its application reduces $\Delta$ of the Structure component.

We observe that \microSSIM, which has background removal, downscaling, and $\alpha$ scaling, consistently gets lower $\Delta$. 
This makes \microSSIM more sensitive to the two input images. 
This enables it to give relatively high SSIM values when comparing similar images (row 1) and low SSIM values when comparing dissimilar images (row 2) simultaneously. 
\paragraph{\textbf{Importance of dataset level estimation of $\beta$}} 
We discover that one of the reasons \microSSIM gives low SSIM value when comparing GT with a pure noise image $I_n$ is that we subtract out $\beta$ from $I_n$, where $\beta$ was estimated from denoised predictions (and not from $I_n$).
This causes the luminance component, and therefore the SSIM score to be low.
If one were to estimate $\beta$ separately for each image pair, as is done in CARE-SSIM, we would not get a low SSIM value.
For SSIM, we show this in Fig.~\ref{fig:saturation_ablation_mito} and \supfigure{\ref{fig:saturation_ablation_actin}} where the SSIM variant \textit{Inst-bkg-rm} has high SSIM value when comparing noise with GT.
For \microSSIM, we show this quantitatively in \suptable{~\ref{tab:datasetlevelparams}}.
This justifies our choice of using a single set of transformation parameters for the entire dataset. 
\paragraph{\textbf{Inspecting role of offset ($\beta$) on vanilla SSIM}}
We model the offset added by the microscope with a single scalar value added to both GT and the prediction.
In Fig.~\ref{fig:luminance_and_uniqueness_ablation}, we investigate the case when the same offset is present in both high-SNR and low-SNR images.
For this experiment, we randomly pick a GT and the corresponding N2V's prediction pair and normalize them as is done for \microSSIM. 
We then create multiple (prediction, GT) pairs by adding a specific offset to the original GT and prediction. 
We compute the SSIM and its components for every offset and plot the curves. 
As can be seen from Fig.~\ref{fig:luminance_and_uniqueness_ablation}, the offset changes the luminance, and therefore the SSIM value.
This observation is important because this means that the SSIM value, and therefore, the quantification of a denoiser's performance, depends upon the offset added by the detector. 
This is certainly undesirable.
Our pre-processing operation of background subtraction removes this offset and therefore removes this unwanted dependence on the SSIM value.
While it is common to normalize by removing the mean, results in Fig.~\ref{fig:bkg_removal_ablation} and (CARE-SSIM results) in Fig.~\ref{fig:baseline_comparisons} highlights issues with mean-removal.
\paragraph{\textbf{Inspecting uniqueness of $\alpha$}}
In Fig.~\ref{fig:luminance_and_uniqueness_ablation} (right), we do a grid search and vary the $\alpha$ from $0$ to $300$. 
As can be seen from the plot, a unique scaling factor exists that yields the optimal SSIM value.
This observation enabled us to use an off-the-shelf optimizer to estimate this scaling factor.
\paragraph{\textbf{Extension to other SSIM variants}}
Since our proposal involves pre-processing followed by estimating the scalar factor $\alpha$, it can be easily incorporated into other popular SSIM variants. 
To this end, we extend MS-SSIM to \MicroMSSSIM. 
We present \microSSIM and \MicroMSSSIM results of unsupervised denoising by N2V and of joint splitting with unsupervised denoising by \denoiSplit in \suptable{\ref{tab:n2v_performance}} and \suptable{\ref{tab:splitting}} respectively.
In \supfigure{\ref{fig:msssimIssues}}, we show that MS-SSIM also has saturation issues and that our \MicroMSSSIM exhibits correct behavior.
\section{Conclusion}
In this work, we propose \microSSIM, a new variant of SSIM that improves the behavior of SSIM on microscopy data.
We explored different unique aspects of the microscopy data namely intensity mismatch between low-SNR and high-SNR micrographs, large intensity pixel values, and an arbitrary digitally added offset. 
We discovered how these peculiarities caused SSIM to behave unexpectedly and formulated a quantifiable notion of saturation to explain the behavior.
We came up with \microSSIM together with an appropriate pre-processing that addressed those. 
We showed empirically that \microSSIM focuses on foreground regions and compared to SSIM, it is more sensitive to the inputs, \ie, worse predictions lead to a larger decrement in the score. 
With \MicroMSSSIM, we show that our approach can be easily extended to other SSIM variants. Finally, we discuss the limitations of our approach in~\supsection{\ref{sec:limitations}}.

In general, we believe that microscopy data is different from natural images in several aspects and more focus should go into understanding these differences and adapting the methods and measures developed for natural images for them to be used on microscopy data. 
Our work is one such effort in this direction.

\paragraph{Acknowledgements}
This work was supported by 
the European Commission through the Horizon Europe program (IMAGINE project, grant agreement 101094250-IMAGINE and AI4LIFE project, grant agreement 101057970-AI4LIFE).
%
%
\bibliographystyle{splncs04}
\bibliography{main}
\newpage

\begin{center}
  \textbf{\Large Supplement \\
  MicroSSIM: Improved Structured Similarity for Comparing Microscopy Data}\\[.2cm]
  
\end{center}
\setcounter{equation}{0}
\setcounter{figure}{0}
\setcounter{table}{0}
\setcounter{page}{1}
\renewcommand{\thepage}{S\arabic{page}} 
\renewcommand{\thefigure}{S\arabic{figure}}
\renewcommand{\thetable}{S\arabic{table}}
\renewcommand*{\thesection}{S\arabic{section}}
\setcounter{section}{0}

 
%

\section{Limitations and future work}
\label{sec:limitations}
One of the limitations of \microSSIM is our assumption that the background regions in the micrographs mostly hover around a low pixel value.
While this is indeed the case often, it is not rare to find our-of-focus fluorescence, bleed-through, or other sources leading to significant pixel intensities in the background region.
In such cases, the optimization for $\alpha$ will suffer because the mentioned sources will typically be present in only either the low-SNR or the corresponding high-SNR micrographs.

We argue that saturation is the critical reason that causes high SSIM score between a micrograph and an image containing pure noise.
While we showed that \microSSIM has lower saturation than other SSIM variants, \microSSIM does not explicitly has any formulation to control saturation.
So, one avenue for improvement is to develop better optimization procedures that can more effectively limit saturation.
A different direction can be to find systematic ways to obtain the constants $c_1,c_2$, and $c_3$ with the objective of having low saturation. 
\section{Derivation for optimality of $\alpha$}
\label{sec:deriveoptimalalpha}
Below, we present the equation for the derivative of SSIM with respect to $\alpha$.
\begin{equation}
    \begin{split}
    \dfrac{\mathrm{d}}{\mathrm{d}\alpha} \text{SSIM}(x, \alpha y)  = - \frac{2 \alpha s_{y}^{2} \cdot \left(2 \alpha s_{xy} + c_{2}\right) \left(2 \alpha u_{x} u_{y} + c_{1}\right)}{\left(\alpha^{2} s_{y}^{2} + c_{2} +
    s_{x}^{2}\right)^{2} \left(\alpha^{2} u_{y}^{2} + c_{1} + u_{x}^{2}\right)} \\
    - \frac{2 \alpha u_{y}^{2} \cdot \left(2 \alpha s_{xy} + c_{2}\right) \left(2 \alpha u_{x}
    u_{y} + c_{1}\right)}{\left(\alpha^{2} s_{y}^{2} + c_{2} + s_{x}^{2}\right) \left(\alpha^{2} u_{y}^{2} + c_{1} + u_{x}^{2}\right)^{2}} \\
    + \frac{2 s_{xy} \left(2 \alpha u_{x} u_{y} + c_{1}\right)}{\left(\alpha^{2} s_{y}^{2} + c_{2} + s_{x}^{2}\right) \left(\alpha^{2} u_{y}^{2} + c_{1} + u_{x}^{2}\right)} \\ 
    + \frac{2 u_{x} u_{y} \left(2 \alpha s_{xy} + c_{2}\right)}{\left(\alpha^{2} s_{y}^{2} + c_{2} + s_{x}^{2}\right) \left(\alpha^{2} u_{y}^{2} + c_{1} + u_{x}^{2}\right)}
    \end{split}
\label{eq:ssimderivative}
\end{equation}
As mentioned in the main paper, existing numerical solvers are unable to find a closed-form solution for $\dfrac{\mathrm{d}}{\mathrm{d}\alpha} \text{SSIM}(x, \alpha y) = 0$. 
However, if we set $c_1=c_2=0$, then the expression for $\text{SSIM}(x,\alpha y)$ simplifies to 
\begin{equation}
    \frac{4 \alpha^{2} s_{xy} u_{x} u_{y}}{\left(\alpha^{2} s_{y}^{2} + s_{x}^{2}\right) \left(\alpha^{2} u_{y}^{2} + u_{x}^{2}\right)},
\end{equation}
and setting its derivative to 0 leads to the following equation
\begin{equation}
    \alpha^4 = \frac{\mu_x^2s_x^2}{\mu_y^2s_y^2}
\end{equation}
Assuming $\mu_x$,$\mu_y$, and $s_{xy}$ to be positive and using the constraint that $\alpha$ should be real and positive, we get the unique expression for $\alpha$ presented in Eq.~\ref{eq:optimalalpha} in the main paper for which the derivative is zero. The double derivative expression $\frac{d^2}{d\alpha^2}(\text{SSIM}(x, \alpha y))$, evaluated on this value of $\alpha$ is
\begin{equation}
    - \frac{32 s_{y}^{2} s_{xy} u_{x} u_{y}^{3}}{s_{x}^{4} u_{y}^{4} + 4 s_{x}^{3} s_{y} u_{x} u_{y}^{3} + 6 s_{x}^{2} s_{y}^{2} u_{x}^{2} u_{y}^{2} + 4 s_{x} s_{y}^{3} u_{x}^{3} u_{y} + s_{y}^{4} u_{x}^{4}}.
\end{equation}
With all the variables involved in the above expression being positive, we can see that the double derivative is negative and therefore we have maxima at $\alpha$'s value proposed in Eq.~\ref{eq:optimalalpha}.
Note that $s_{xy}\approx0$ when comparing two totally uncorrelated images (for eg., random noise). For any reasonable prediction y, it is the case that $s_{xy}>0$ where x is the ground truth. So, our assumption of $s_{xy} >0$ is justified. 

\begin{table}[]
    \centering
    \begin{tabular}{c|c|c|c|c|c|c}
          \multirow{2}{*}{Denoising-Task} & \multicolumn{3}{c|}{Saturation $\Delta$} &\multirow{2}{*}{\microSSIM} & \multirow{2}{*}{\MicroMSSSIM}\\\cline{2-4}
         &$l$ &$c$ & $s$ & &   \\
         \hline
         Actin &4.2\scriptsize{ $\pm$ 3.1}&143\scriptsize{ $\pm$ 114}&0.7\scriptsize{ $\pm$ 0.7}& 0.68\scriptsize{ $\pm$ 0.086} & 0.83\scriptsize{ $\pm$ 0.055}\\
         Mitochondria &7.1\scriptsize{ $\pm$ 23.4} & 256\scriptsize{ $\pm$ 925} & 0.1\scriptsize{ $\pm$ 0.2} & 0.84\scriptsize{ $\pm$ 0.051} & 0.910\scriptsize{ $\pm$ 0.034} \\
    \end{tabular}
    \caption{Denoising performance by N2V\cite{n2v-krull}. N2V was separately trained on Actin and Mitochondria low-SNR micrographs. We denoise all the 100 $2048\times 2048$ sized low-SNR micrographs and compare them against the corresponding high-SNR micrographs, also provided in the dataset. We also report the saturation factor $\Delta$, mean and standard deviation over the 100 micrographs, for each of the three SSIM components.}
    \label{tab:n2v_performance}
\end{table}
\begin{table}[]
    \centering
    \begin{tabular}{c|c|c}
         Channel & \microSSIM & \MicroMSSSIM \\
         \hline
         Actin & 0.68\scriptsize{ $\pm$ 0.049} & 0.830\scriptsize{ $\pm$0.044} \\
         Mitochondria & 0.81\scriptsize{ $\pm$0.059} & 0.90\scriptsize{ $\pm$0.038} \\
    \end{tabular}
    \caption{Joint splitting and unsupervised denoising performance by \denoiSplit~\cite{denoisplit}. We trained the \denoiSplit model on the Actin vs Mitochondria task using the low-SNR data provided in the Hagen et al.~\cite{Hagen2021-xh}.}
    \label{tab:splitting}
\end{table}
\begin{table}[]
    \centering
    \begin{tabular}{c|c}
        Instance level & Dataset level  \\
        \hline
         0.61\scriptsize{$ \pm $0.131}& -0.17\scriptsize{$ \pm $0.087} \\  
    \end{tabular}
    \caption{We selected 30 full-frame high-SNR GT from Actin denoising task. We computed \microSSIM between these images and the pure uniform noise in two ways. 
    In column \textit{Dataset level}, we compute \microSSIM in the way proposed in this work. In column \textit{Instance level}, we estimate $\beta$, the background, for each pure noise instance separately and use that in the pre-processing step. Note that in this case, for background regions of the image, both pre-processed random noise and GT will have the mean pixel value hovering around zero and so the luminance component becomes high. This leads to an undesirably higher SSIM value. }
    \label{tab:datasetlevelparams}
\end{table}
\clearpage
\section{Optimization Configuration}
\label{sec:optimizationhparams}
We used $\text{scipy}==1.11.3$ python package for optimization. All optional parameters were left to their default values. Below we provide a snippet from our code for clarification 
\begin{lstlisting}
    from scipy.optimize import minimize
    ...
    ...
    def get_ssim(alpha, ux, uy, vx, vy, vxy, C1, C2, C3=None):
        "Returns SSIM score"
        ...
    ...
    
    other_args = (ux,uy,vx,vy,vxy,C1,C2,)
    initial_guess = np.array([1])
    res = minimize(
        lambda *args: -1 * get_ssim(*args), initial_guess, 
        args=other_args, method=None, jac=None, 
        hess=None,hessp=None, bounds=None, 
        constraints=(), tol=None, 
        callback=None, options=None)
    return res.x[0]
\end{lstlisting}
\figMSSSIMissues
\figAblationSaturationActin

\begin{figure}[]
\centering
\includegraphics[width=.6\textwidth]{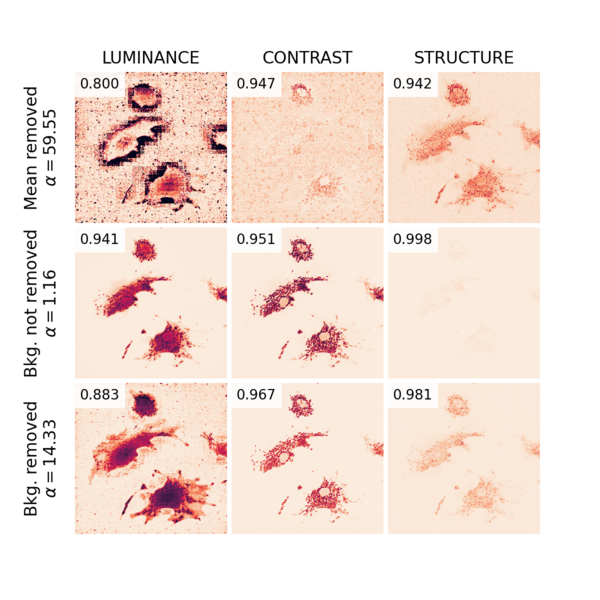}
\caption{Ablation on background removal on Mitochondria denoising task. Individual grids are heatmaps and the colorbar used is identical to what is used in the main manuscript Fig.~\ref{fig:bkg_removal_ablation}.}
\label{fig:bkg_removal_ablation0}
\end{figure}

\begin{figure}[]
\centering
\includegraphics[width=.6\textwidth]{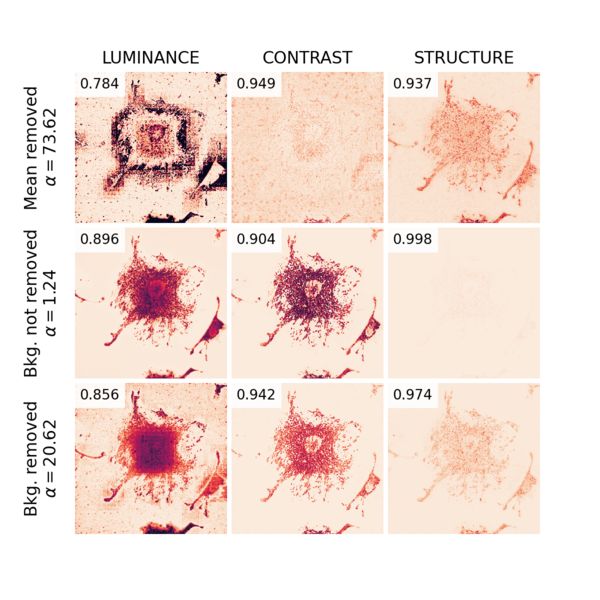}
\caption{Ablation on background removal on Mitochondria denoising task. Individual grids are heatmaps and the colorbar used is identical to what is used in the main manuscript Fig.~\ref{fig:bkg_removal_ablation}.}
\label{fig:bkg_removal_ablation10}
\end{figure}

\begin{figure}[]
\centering
\includegraphics[width=.6\textwidth]{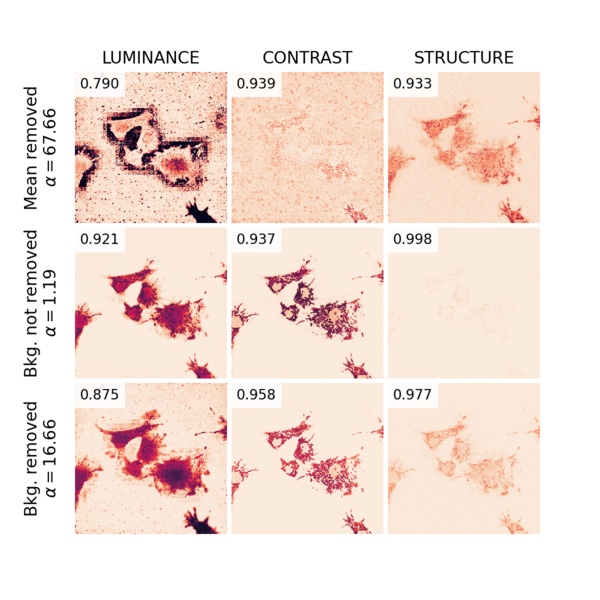}
\caption{Ablation on background removal on Mitochondria denoising task. Individual grids are heatmaps and the colorbar used is identical to what is used in the main manuscript Fig.~\ref{fig:bkg_removal_ablation}.}
\label{fig:bkg_removal_ablation20}
\end{figure}

\begin{figure}[]
\centering
\includegraphics[width=.6\textwidth]{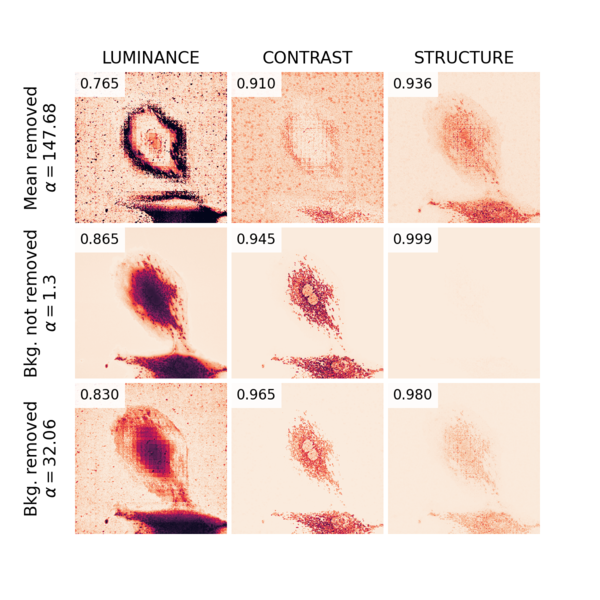}
\caption{Ablation on background removal on Mitochondria denoising task. Individual grids are heatmaps and the colorbar used is identical to what is used in the main manuscript Fig.~\ref{fig:bkg_removal_ablation}.}
\label{fig:bkg_removal_ablation30}
\end{figure}

\begin{figure}[]
\centering
\includegraphics[width=.6\textwidth]{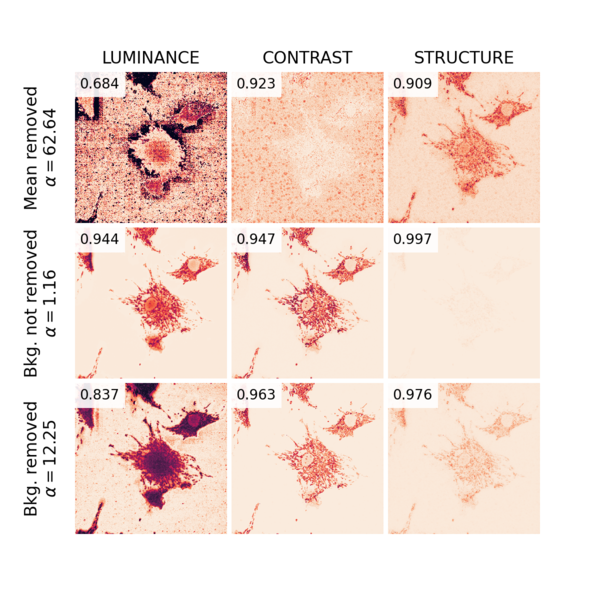}
\caption{Ablation on background removal on Mitochondria denoising task. Individual grids are heatmaps and the colorbar used is identical to what is used in the main manuscript Fig.~\ref{fig:bkg_removal_ablation}.}
\label{fig:bkg_removal_ablation40}
\end{figure}

\begin{figure}[]
\centering
\includegraphics[width=.6\textwidth]{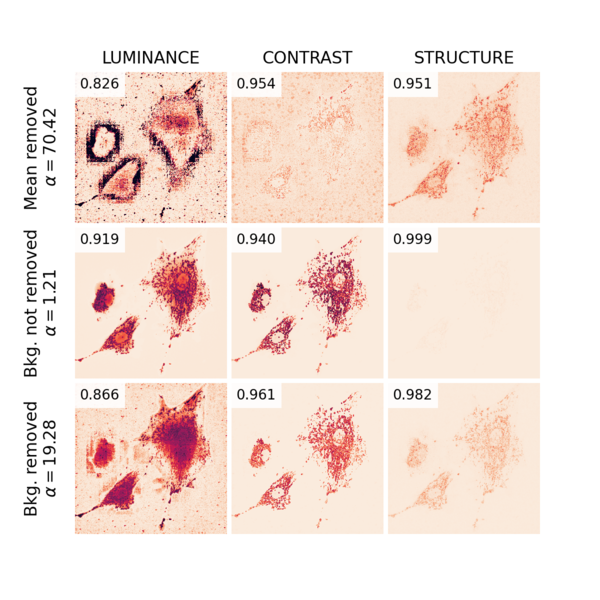}
\caption{Ablation on background removal on Mitochondria denoising task. Individual grids are heatmaps and the colorbar used is identical to what is used in the main manuscript Fig.~\ref{fig:bkg_removal_ablation}.}
\label{fig:bkg_removal_ablation50}
\end{figure}

\begin{figure}[]
\centering
\includegraphics[width=.6\textwidth]{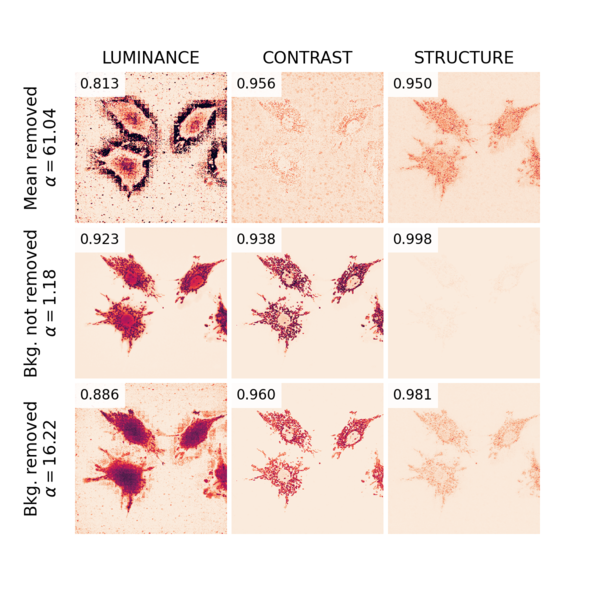}
\caption{Ablation on background removal on Mitochondria denoising task. Individual grids are heatmaps and the colorbar used is identical to what is used in the main manuscript Fig.~\ref{fig:bkg_removal_ablation}.}
\label{fig:bkg_removal_ablation60}
\end{figure}

\begin{figure}[]
\centering
\includegraphics[width=.6\textwidth]{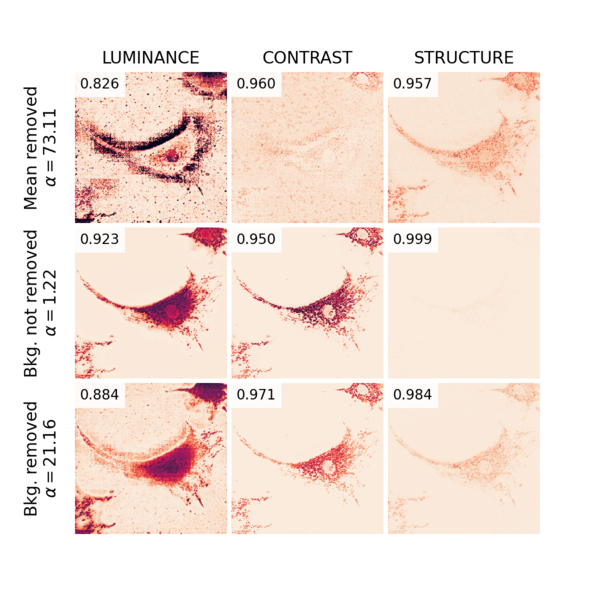}
\caption{Ablation on background removal on Mitochondria denoising task. Individual grids are heatmaps and the colorbar used is identical to what is used in the main manuscript Fig.~\ref{fig:bkg_removal_ablation}.}
\label{fig:bkg_removal_ablation70}
\end{figure}
%
%
%
%
%
%
\begin{figure}[]
\centering
\includegraphics[width=.6\textwidth]{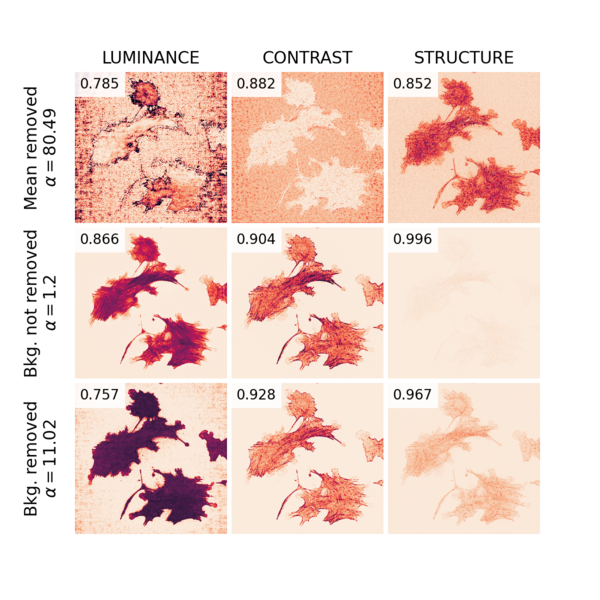}
\caption{Ablation on background removal on Actin denoising task. Individual grids are heatmaps and the colorbar used is identical to what is used in the main manuscript Fig.~\ref{fig:bkg_removal_ablation}.}
\label{fig:bkg_removal_ablation0}
\end{figure}

\begin{figure}[]
\centering
\includegraphics[width=.6\textwidth]{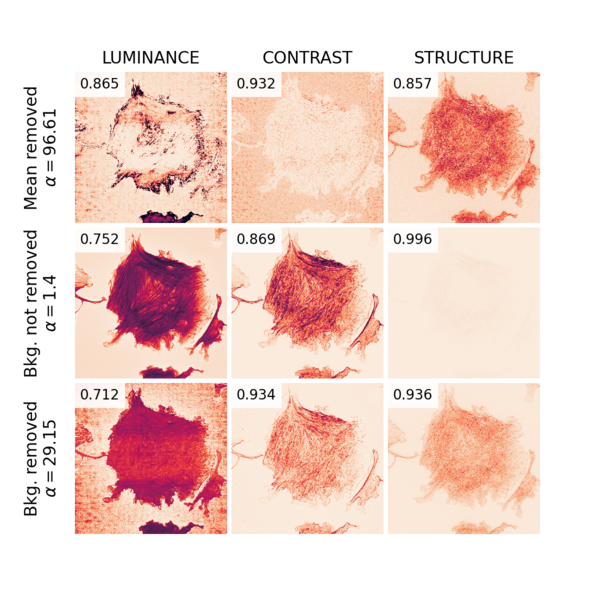}
\caption{Ablation on background removal on Actin denoising task. Individual grids are heatmaps and the colorbar used is identical to what is used in the main manuscript Fig.~\ref{fig:bkg_removal_ablation}.}
\label{fig:bkg_removal_ablation10}
\end{figure}

\begin{figure}[]
\centering
\includegraphics[width=.6\textwidth]{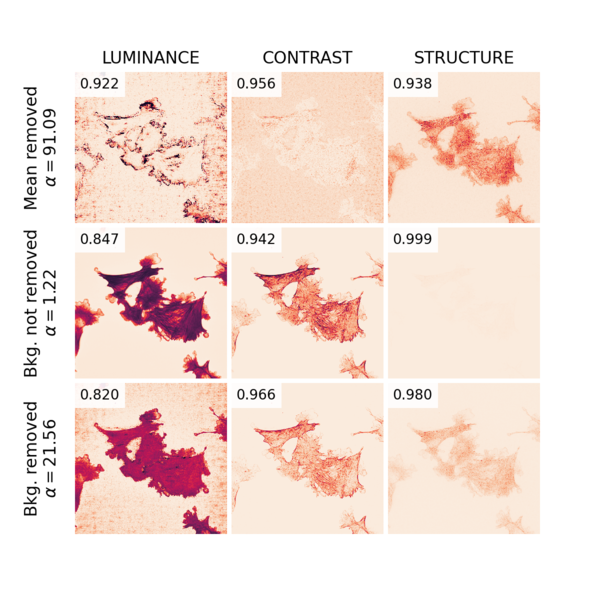}
\caption{Ablation on background removal on Actin denoising task. Individual grids are heatmaps and the colorbar used is identical to what is used in the main manuscript Fig.~\ref{fig:bkg_removal_ablation}.}
\label{fig:bkg_removal_ablation20}
\end{figure}

\begin{figure}[]
\centering
\includegraphics[width=.6\textwidth]{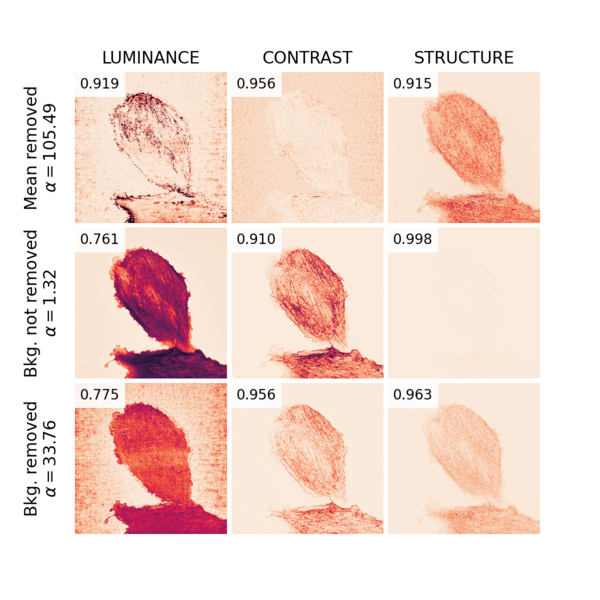}
\caption{Ablation on background removal on Actin denoising task. Individual grids are heatmaps and the colorbar used is identical to what is used in the main manuscript Fig.~\ref{fig:bkg_removal_ablation}.}
\label{fig:bkg_removal_ablation30}
\end{figure}

\begin{figure}[]
\centering
\includegraphics[width=.6\textwidth]{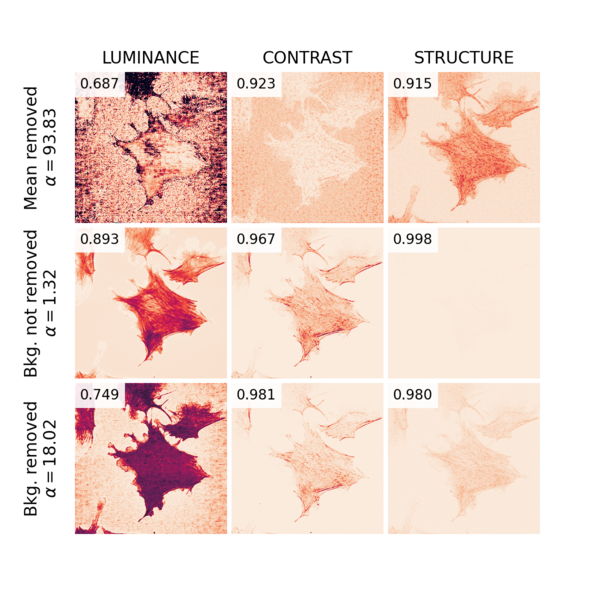}
\caption{Ablation on background removal on Actin denoising task. Individual grids are heatmaps and the colorbar used is identical to what is used in the main manuscript Fig.~\ref{fig:bkg_removal_ablation}.}
\label{fig:bkg_removal_ablation40}
\end{figure}

\begin{figure}[]
\centering
\includegraphics[width=.6\textwidth]{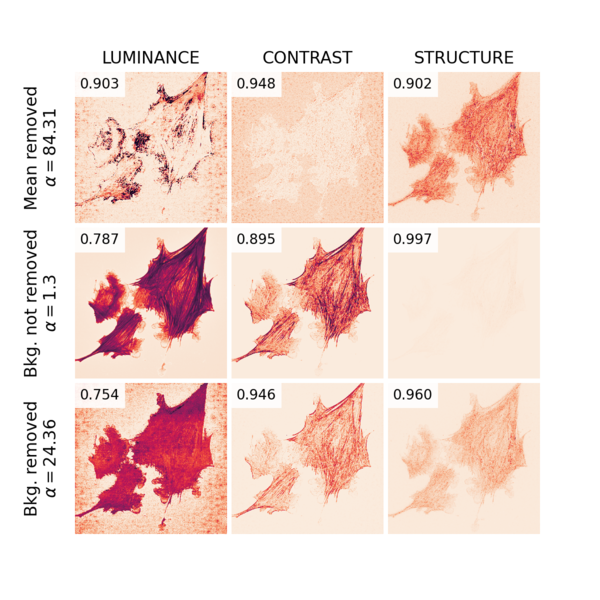}
\caption{Ablation on background removal on Actin denoising task. Individual grids are heatmaps and the colorbar used is identical to what is used in the main manuscript Fig.~\ref{fig:bkg_removal_ablation}.}
\label{fig:bkg_removal_ablation50}
\end{figure}

\begin{figure}[]
\centering
\includegraphics[width=.6\textwidth]{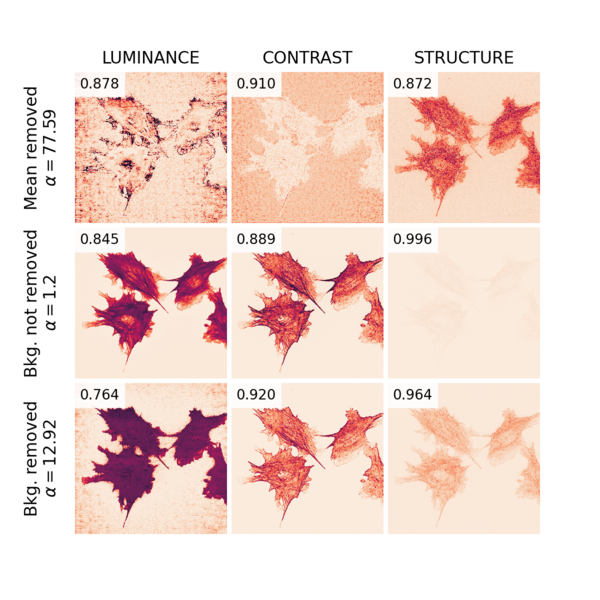}
\caption{Ablation on background removal on Actin denoising task. Individual grids are heatmaps and the colorbar used is identical to what is used in the main manuscript Fig.~\ref{fig:bkg_removal_ablation}.}
\label{fig:bkg_removal_ablation60}
\end{figure}

\begin{figure}[]
\centering
\includegraphics[width=.6\textwidth]{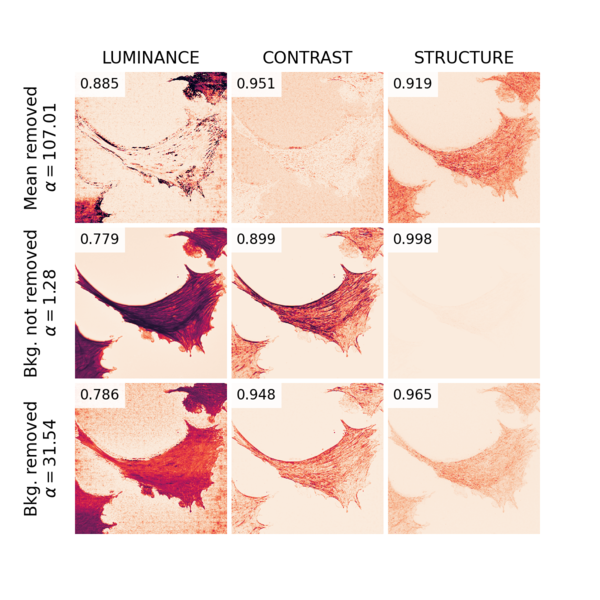}
\caption{Ablation on background removal on Actin denoising task. Individual grids are heatmaps and the colorbar used is identical to what is used in the main manuscript Fig.~\ref{fig:bkg_removal_ablation}.}
\label{fig:bkg_removal_ablation70}
\end{figure}

\end{document}